\documentclass[referee]{cjaa}
\usepackage{natbib}
\usepackage{psfig}
\def\mnras{Mont. Not. R. Astron. Soc.}
\def\apj{ApJ}
\def\apjl{ApJ}

\def\nat{Nature}
\def\aap{A\&A} 
\def\aaps{A\&A Supplements }

\begin{document}

\title{ 
On the Large-Scale Structure of the Universe
as given by the Voronoi Diagrams 
  }
   \volnopage{Vol.0 (200x) No.0, 000--000}      %%preserved for Editor. DOn't remove!
   \setcounter{page}{1}          %%starting page, preserved for Editor. DOn't remove!

\author    {L. Zaninetti  \inst {1}}
\offprints {L. Zaninetti}
\institute {Dipartimento di Fisica Generale, \\
           Via Pietro Giuria 1               \\
           10125 Torino, Italy               \\
  \email   { zaninetti@ph.unito.it}
}
\date      {Received \dotfill ; accepted \dotfill}

\abstract{
The size distributions of 2D and 3D Voronoi cells and of 
cells of $V_p(2,3)$,---2D cut of 3D Voronoi diagram---are explored,
with the single-parameter (re-scaled) gamma distribution playing
a central role in the analytical fitting. 
Observational evidence for a cellular universe is briefly reviewed. 
A simulated $V_p(2,3)$ map with galaxies lying on the 
cell boundaries is constructed to compare, 
as regards general appearance,
with the observed CFA map of galaxies and voids, 
the parameters of the simulation being so chosen as to 
reproduce the largest observed void size.
}
\authorrunning {L. Zaninetti}
\titlerunning  {Cosmic Voids}

\maketitle  

\section{Introduction}

The applications of the Voronoi diagrams (see \citet {voronoi}) in
astrophysics started with ~\citet{kiang} where the size
distribution in 1D as given by random seeds was theoretically
deduced  in  a rigorous way. \citet{kiang} also derived~, performing
a Monte Carlo experiment,   the area distribution in 2D and volume
distribution in 3D. The idea that area and volume distributions
follow a gamma--variate with argument 4 and 6 respectively  was
later reported as "Kiang's conjecture", see~\citet{okabe}. 
The
application of  the Voronoi diagrams to the distribution of
galaxies started with~\citet{icke}, where a sequential clustering
process was adopted in order to insert the initial seeds. Later
on~\citet{pierre} introduced a  general algorithm for simulating
one-dimensional lines of sight through a cellular universe~. The
large microwave background temperature anisotropies over angular
scales up to one degree were calculated using a Voronoi model for
large-scale structure formation in \citet{barrow} and
\citet{coles1991}.

The possibility to explain the CFA slices using a fractal
distribution of seeds  and inserting the galaxies on the faces of
the irregular polyhedron was explored by~\citet{zaninettig}~.

A detailed Monte Carlo simulation of pencil beam-like redshift
surveys  was carried out by ~\citet{subba}: they found
 that the probability of  finding regularity varies
from 3 to 15 percent depending on the details of the models.

Another Monte Carlo study  was carried out  by ~\citet{van}
where three different distributions of nuclei were adopted in
order to perform extensive statistical analysis of several
geometrical aspects of three dimensional Voronoi tessellation.
A new way of partitioning 
the space
into cells characterised by the shape of rhombic dodecahedron
has 
been introduced in~\citet{kiang2003} ;
the  application is done to the  CfA catalogue 
and to the   IRAS/PSCz catalogue~\citet{kiang2004}.
The void hierarchy approach 
has been introduced in~\citet{sheth} 
and it explains  how large-scale
structures are function of two parameters, 
one of which reflects the dynamics of void formation, 
and the other the
formation of collapsed objects. 

From the point of view of the astronomical  observations a few
papers that  point toward the cellular  structure of the universe
are now briefly reviewed.
On analysing the data  from four distinct surveys at the north and
south Galactic poles ~\citet{broadhurst} found an apparent
regularity in the galaxy distribution with a characteristic scale
of 128  Mpc. The astronomers that analysed 
 the maps of the galaxy distribution up to
cz=15000 Km/s , see for example 
\citet{delapparent} and  \citet{geller}, found  large coherent structures: the
largest void found having a diameter of 5000 km/s.

Great advances  in the observational data 
, see  \citet{folkes} , \citet{ratcliffe} and  \citet{shectman},
brought the limits of the observations at cz=60000 Km/s and
confirmed the existence of voids in the distribution of galaxies.

The distribution of clusters in rich super-clusters is
not isotropic: it is periodic along a cubic lattice approximately
aligned with the super-galactic coordinates, see for
example~\citet{saar}.

The Voronoi diagrams  are also  used to process the astronomical
data , see ~\citet {elad} and \citet{ramella} .
As an example  \citet{ramella} implemented a  Voronoi Galaxy
Cluster Finder that  uses galaxy positions and magnitudes to find
clusters and determine their main features: size, richness and
contrast above the background.

The starting point  is to consider a series of explosions that
 start  at the same time in a homogeneous space. The shells
connected with the explosions  meet on a 3D network given by the
nested irregular polyhedron. 
From an astrophysical point of view 
this network can be realized by
a set of primordial explosions, see~\citet{charlton} and \citet{ferraro}, 
described by the Sedov solution 
in the adiabatic phase:
\begin{equation}
R(t)=
\bigl ({\frac {25}{4}}\,{\frac {{\it E}\,{t}^{2}}{\pi \,\rho}}\bigr)^{1/5}
=
12.49 Mpc\,\bigl({{\frac {{\it E_{64}}\,{{\it t_9}}^{2}}{{\it n_{-7}}}}}
\bigr )^{1/5}
\label{rprimeval}
\quad ,
\end{equation}
where  t represents the time                  , 
E is the energy injected in the explosion     ,
$\rho$ is the density of matter               ,
$\rho=n$m                                     , 
n is number of particles per unit volume      ,
m=1.4$m_{\mathrm {H}}$                        ,
$m_{\mathrm {H}}$ is the mass of the hydrogen ,
$t_9=\frac {t}{10^9~yr}$                      ,
$E_{64}=\frac{E}{10^{64}~erg}$ 
and     $n_{-7}=\frac{n}{10^{-7}{\mathrm{particles~}}{\mathrm{cm}^{-3}}}$ .

The already cited works leave the following  questions unanswered
or partially answered.

Is the "Kiang's conjecture" applicable in  2D and 3D environment
with physical parameters near to those of the galaxies~?

What is the  probability density function that characterises  the
two--dimensional sectional area connected with 3D Voronoi
diagrams~?

Can the  averaged area connected with 
the voids in the distribution of galaxies
 visible on the
CFA2 slices be guessed  from the theory~?

Can the number of theoretical
voids in the distribution of galaxies
 in a sphere of radius equal to
that of the CFA2 slices  be deduced theoretically~?

In order  to answer these questions 
the coefficient of the
gamma--variate that characterises the area--distribution  of a 2D
Voronoi diagram and the volume--distribution  in 3D were derived
in \S~\ref{area2D_section} and in \S~\ref{volume}
respectively.

 The coefficient of the gamma--variate
 that characterises the sectional area of a 3D Voronoi diagram
 as well some characteristics of the 
voids in the distribution of galaxies
 were  derived in
 \S~\ref{cut2D}.

The observed large scale structures of galaxies are classified
as CFA slices , LCRS slices or pencil beam surveys :
they are simulated in \S~\ref{galaxies}.

The number of seeds necessary to produce a theoretical framework
comparable to the observed one was computed in
\S~\ref{voidsdensity}~.

\section{The preliminaries}

The type of adopted lattice , the importance of setting 
properly the boundary conditions, the type  of seeds that 
generates the polygons/polyhedron , the concept of unitarian 
area and volume  and a first two dimensional scan are now introduced.

\subsection{The adopted lattice}

Our  method considers a 2D and a  3D lattice
made of  ${\it pixels}^{2}$  and   ${\it pixels}^{3}$ points~:
present in this lattice are $N_s$ seeds generated
according to a random process.
All the computations are usually performed on this mathematical
lattice; 
the conversion to the physical lattice
is obtained   by multiplying the unit
by $\delta=\frac{side}{pixels -1}$ , where {\it side}
is the  length of the square/cube expressed in the physical
unit  adopted.

\subsection{Boundary conditions}

In order to minimise boundary  effects introduced by those
polygons/polyhedron that cross the square/cubic boundary,
 we amplify
the area/cube in which the seeds are inserted by  a factor {\it
amplify}~. Therefore the N seeds are inserted in an  area/volume
that is $  pixels^2 \times amplify$  or $  pixels^3 \times
amplify$ ,  
which is bigger than  the box over which we perform the
scanning; {\it amplify } is generally taken to be equal to 1.2~.
This procedure inserts periodic boundary conditions to our
square/cube. 
The number of seeds that fall  in the area/cube is
$N_s$ with  $N_s < N$. In order to avoid  computing 
 incomplete  area/volumes  we select the cells 
that do not intersect the square/cubic
boundary. This is  obtained by selecting the cells that
belong to seeds which are comprised in an area/volume that is {\it
select} times smaller than  $  pixels^2 $ or $  pixels^3 $; of
course {\it select} is smaller than one and varies between 0.1
and 0.5.

\subsection{The seeds }

The points are generated independently on the X and Y axis in 2D (
adding the Z axis in 3D ) through a subroutine  that returns a
pseudo-random real number taken from a uniform distribution
between 0 and 1~. For practical purposes we used the subroutine
RAN1 as described in~\citet{press}~.

\subsection{The adopted units  }

In order to deal with quantities of the order of one we divide the
obtained area/volume in units  of  ${\it pixels}^2/{\it pixels}^3$
by the  expected unitarian area/volume  , $u_A$ and $u_V$
respectively 
\begin{eqnarray*}
u_A = {\it pixels }^2 \times {\it amplify} / N  \\
u_V = {\it pixels }^3 \times {\it amplify} / N
\quad .
\label{aree_normalizzate}
\end{eqnarray*}
This operation represents a first normalisation.
The expected unitarian  quantities can also be expressed in
physical units, $u_A^p$ and $u_V^p$
\begin{eqnarray*}
u_A^p = {\it side }^2 \times {\it amplify} / N  \\
u_V^p = {\it  side }^3 \times {\it amplify} / N
\quad .
\label{aree_normalizzate_physical}
\end{eqnarray*}
Special attention should be paid
when the area is  computed like
a cut of a 3D network;
this  case is named  $V_p(2,3)$ , see Sec.\ref{vp23}~.
The unitarian area
$u_{A(2,3)}$   is expected
to be
\begin{equation}
u_{A(2,3)}  =
u_V ^{2/3}
\quad  ,
\end{equation}
and the physical counterpart
\begin{equation}
u^p_{A(2,3)}  =
(u^p_V)^{2/3}
\quad  .
\label{area_fisica}
\end {equation}

\subsection{The two dimensional scan }

\label{area2D_section}
\begin{figure}
\begin{center}
\psfig{file=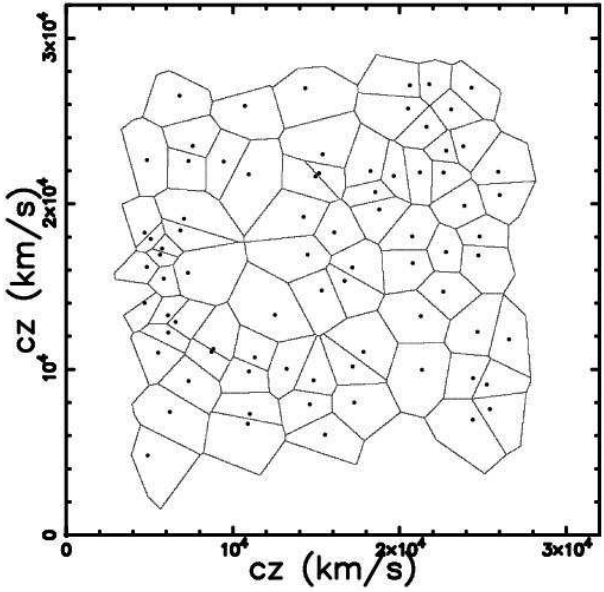,width=8cm} 
\end{center} 
\caption {
The Voronoi--diagram in 2D when random seeds are used.
The  parameters
are {\it pixels}~= 800
       , {\it N}~= 180
  ,{\it amplify}~= 1.2
  ,   {\it side}~= 2 $\times$ 16000 Km/sec
  , {\it select}~= 0.5~ }
          \label{area2D}%
    \end{figure}

\begin{figure}
\begin{center}
\psfig{file=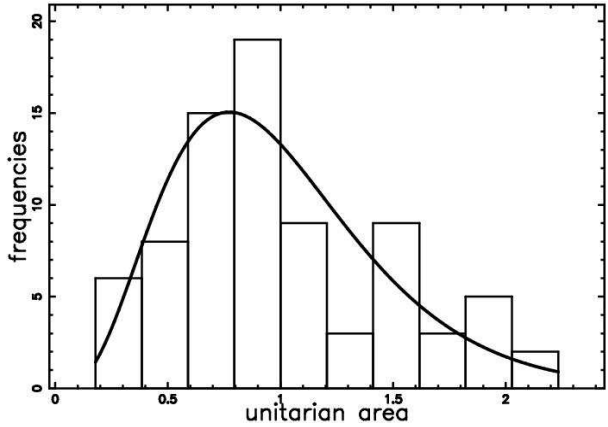,width=8cm} 
\end {center}
\caption {
Histogram (step-diagram)  of area distribution  in 2D
with a superposition of the
fitting line (the gamma--variate);
input parameters  as in Figure~\ref{area2D},
c=4.38,
number of bins =10 ,  
$\chi^2$=12.67 ,
$\overline{A}$ = 0.98 ,
$A_{max} $ =  2.19   , 
$ A_{min}$ =  0.17  .
          }
          \label{2D_area_gamma}%
    \end{figure}

A lattice   made of ({\it pixel})$^2$ points is considered
and
a typical run using the seeds as given  by  a random  process  is
visualised   in Figure~\ref{area2D}.
Once the histogram of the area  is obtained, 
we can fit it~,
following ~\cite{kiang}, with the following one parameter 
probability density function , in the following 
pdf:
\begin{equation}
 H (x ;c ) = \frac {c} { \Gamma (c)} (cx )^{c-1} \exp(-cx)
\quad ~,
\label{kiang}
\end{equation}
where $  0 \leq x < \infty $ , $ c~>0$
and  $\Gamma (c)$ is the gamma function with argument c.
This pdf is characterised
by $\mu$=1 and $\sigma^2$=1/c.
The value of c   is obtained from the
method of the matching moments
\begin{equation}
 c = \frac {1}{\sigma^2}=
\frac  {n-1}  { \sum_{i=1}^n  ( x_i - 1  )^2 }
\quad .
\end {equation}
The data should be
normalised  in order  to have ${\overline{x}}$~=1.
The  frequency histogram and
the relative best fit through  the gamma-variate are plotted
 in
Figure~\ref{2D_area_gamma} .
The captions  of Figure~\ref{2D_area_gamma}  report also 
the following quantities expressed in normalised
area units (see
formula~(\ref{aree_normalizzate}~))
$\overline{A}$,  $A_{max} $
, $ A_{min}$ ,respectively the averaged,
the maximum and the minimum of the area--sample
and  $\chi^2$ , that represents
the  goodness of fit.
%begin figure area2D
% end figure area2D
%begin figure 2D_area_gamma
% end figure 2D_area_gamma

The value of c in the case of random seeds 
is connected with the "Kiang's conjecture"
that in 2D  means c=4 ; 
the numerical evaluations give similar values ,
here  c = 4.38  is found.

\section{The 3D case }

In order to make a comparison with the astronomical observations
the tessellation in $\Re^3$ is firstly analysed
through a planar section  and  
the distribution of  volume
is numerically derived.

\subsection{The   2D cut }

\label{cut2D}
We now work on a 3D lattice  L$_{k,m,n}$ of pixels$^3$ elements .
Given a section  of the cube
(characterised , for example, by $k=\frac{pixel}{2}$)
the  various $V_i$ (the volume belonging
to the seed i)
 may or may not cross the  little
cubes belonging to the two dimensional lattice .

\begin{figure}
\begin{center}
\psfig{file=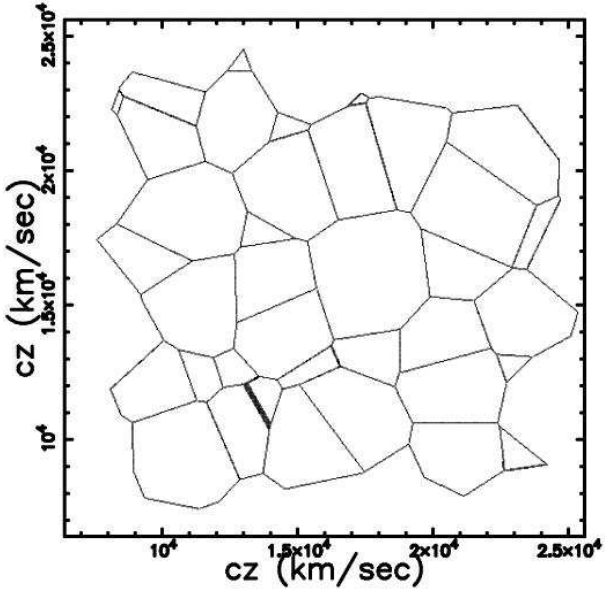,width=8cm} 
\end {center}
\caption {
Portion of the Voronoi--diagram $V_p(2,3)$ when random seeds are used;
cut on the  x--y plane ~.
The  parameters
are {\it pixels}~= 800
       , {\it N}~= 1900
  ,   {\it side}~= 2 $\times$ 16000 Km/sec
  ,{\it amplify}~= 1.2
and {\it select}~= 0.1~.}
          \label{cut_middle}%
    \end{figure}

\begin{figure}
\begin{center}
\psfig{file=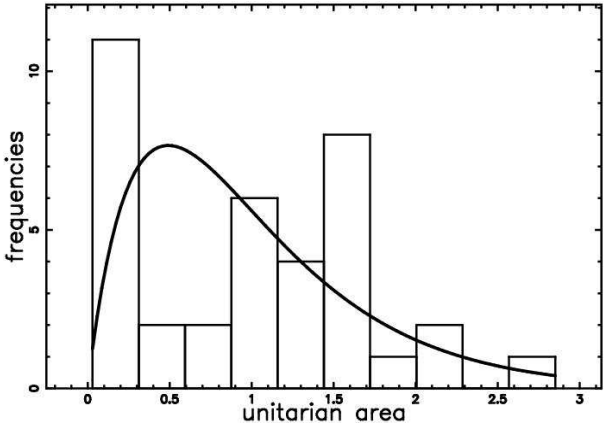,width=8cm} 
\end {center}
\caption {
Histogram (step-diagram)  of $V_p(2,3)$   distribution
on an x-y plane with a superposition of the
fitting line (the gamma--variate);
parameters  as in Figure~\ref{cut_middle},
 c=1.99 $\pm$ 0.22, number of bin =10 , $\chi^2$=23.09 ,
$\overline{A}$ = 0.78 $\pm$ 0.06    ,
$A_{max} $ =  2.12   $\pm$  0.26    , 
$ A_{min}$ =  0.02   $\pm$  0.01    .
}
          \label{3D_area_gamma}%
    \end{figure}

Following the nomenclature introduced by~\citet{okabe} we can call
the intersection between a plane and the cube previously described
as  $V_p(2,3)$. 
\label{vp23} 
A typical result of this 2D sectional
operation in the x--y plane can be visualised in
Figure~\ref{cut_middle}~, the  frequency histogram and the relative
best fit through a gamma-variate  pdf of  the $V_p(2,3)$
distribution are  reported in Figure~\ref{3D_area_gamma}  together
with  the
derived value of c.
Following the hypothesis that the galaxies are distributed 
on the faces of the irregular polyhedra the network of 
Figure~\ref{cut_middle}
represents the spatial coordinates where the galaxies are.
The thick edges of Figure~\ref{cut_middle} represent the intersection 
between the slice and a face ; we remember that the area of intersection
increases with cz.
%begin figure cut_middle
% end figure cut_middle

%begin figure 3D_area_gamma
% end figure 3D_area_gamma
Due to the great importance that the properties of the sample
$V_p(2,3)$ could have in the astrophysical applications
we derived three  samples (all crossing the center)
 of $V_p(2,3)$
on the x-y , x-z and y-z plane.
This allow us to find the average value of the sample properties
and their error, see captions of Figure~\ref{3D_area_gamma}.

The mathematical theory of the expected values  of
characteristics of a typical cell in 1--dimensional sectional
Poisson Voronoi diagram, see ~\citet{okabe}
 , gives
\begin{equation}
\overline{A}=0.68 \lambda ^{-2/3} \quad ,
\end{equation}
where $\lambda$ is the intensity of the Poisson process. In our
case  $\lambda$ is replaced by the
 random points.
 Our values of $\overline{A}$ , see
captions in Figure~\ref{3D_area_gamma}, are near to the values predicted by the
mathematical theory. In order to obtain
the number of edges/crossed~faces
we report in Table~\ref{3D_area_edges} the probability to have n
edges for random seeds  and the theoretical value as
given in~\citet{okabe}.
\begin{table}
 \caption[]{The  probability to have n--edges
            in  $V_p(2,3)$. }
 \begin{tabular}{|c|c|c|c|c|c|c|c|}
 \hline
 \hline
seeds $\setminus$ n  &  3   & 4 & 5 & 6 & 7 & 8  & 9  \\
   \hline
~\citet{okabe}
 &   0.063 &    0.13  & 0.2   &  0.22
& 0.18   &  0.11 & 0.05  \\ \hline
random  &   0.081 &    0.16   & 0.24   &  0.18
& 0.10   &  0.16 & 0.054   \\ \hline
 \hline
 \end{tabular}
 \label  {3D_area_edges}
 \end {table}
The derived values of c remember
the theoretical distribution of 1D Voronoi segments in which c=2,
see~\citet{kiang}~.

\subsection{Projection on the sphere}

\label{sec_sphere}
\begin{figure}
\begin{center}
\psfig{file=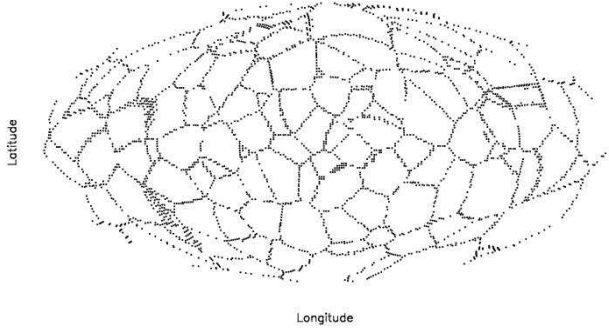,width=8cm} 
\end {center}
\caption {
The Voronoi--diagram $V_p(2,3)$ 
in the Hammer-Aitof  projection
at cz = 7201 Km/sec.}
          \label{aitof_sphere}%
    \end{figure}

Another way to look at the cross sectional area
is a spherical cut characterised by a constant value 
of the distance to the center of the box ,
in this case expressed in cz units, see Figure~\ref{aitof_sphere}~.
%begin figure aitof_sphere
% end figure aitof_sphere

\subsection{The  statistics of the  volume }

\label{volume}

\begin{figure}
\begin{center}
\psfig{file=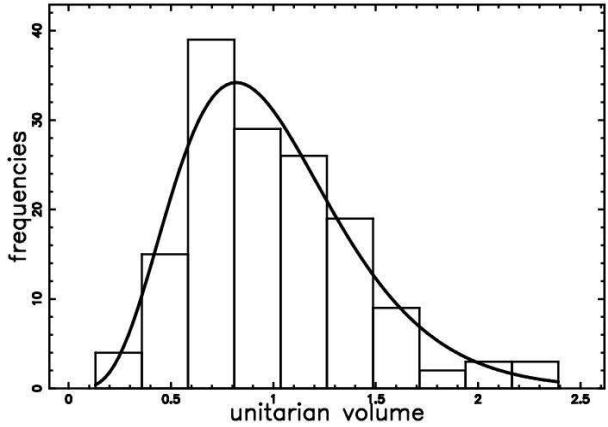,width=8cm} 
\end {center}
\caption {
Histogram (step-diagram)  of volume  distribution
with a superposition of the
fitting line (the gamma--variate).
Parameters  as in Figure~\ref{cut_middle}
but  {\it pixels}~= 400~
: c=5.50, NBIN=10 and $\chi^2$=8.05.}
          \label{3D_volumes_gamma}%
    \end{figure}

In  every  point--lattice L$_{k,m,n}$ 
we compute the nearest 
seed  and 
we increase by one the volume
of that seed.
The  frequency histogram and the relative best fit
through gamma-variate  pdf for the volume distribution
is reported in Figure~\ref{3D_volumes_gamma}.
%begin figure 3D_volumes_gamma
% end figure 3D_volumes_gamma
The experimental frequencies  are
fitted  by a  gamma--variate
with c = 5.5  .
This   value of c should  be compared with the value of 6 as
deduced by \citet{kiang} and successively refined in 5.5  due to a
change  in the generator  of random numbers (~\citet{kiang2})
and  with    5.78 as deduced by \cite{kumar}.

\section{The spatial distribution of galaxies}

\label{galaxies}
The theory of the  sectional area derived in Sec.~\ref{cut2D}
can be the  framework that   explains the existence of voids in
the spatial distribution of galaxies. 
The observational material
that proves the existence of voids in the distribution of galaxies is briefly
reviewed and  then
the number of voids in the distribution of galaxies  that characterises the  CFA2 slices is
derived.
The calibration of  the Voronoi diagrams on the void of largest 
area measurable on the CFA slices   allows to simulate 
the maximum in the  galaxy's concentration present in the various
type of slices  versus 
cz  and the 128 Mpc regularity.

\subsection{The CFA2 slices}

\label{CFA2} 
The second CFA2 redshift   Survey , started in 1984,
produced slices showing that the spatial distribution of galaxies
is not random but distributed on filaments that represent the 2D
projection of 3D bubbles. 
We recall that a slice comprises all the
galaxies with magnitude $m_b~\leq~16.5$ in a strip of $6^{\circ}$
wide and about $130^{\circ}$ long. 
One of such slice (the so
called first CFA strip) is visible at the following address
http://cfa-www.harvard.edu/~huchra/zcat/ ; more details can be
found in~\citet{geller}. 
The already mentioned slice can be
down-loaded from http://cfa-www.harvard.edu/~huchra/zcat/n30.dat/
. 
The first of such slices presents many voids, the bigger one
being 4000$\div$5000 Km/sec across (~\citet{huchra}). The greatest
possible attention should be paid to the derivation of the void
with maximum area ,$A^{obs}_{max}$~,
because is the area that allows  the  determination of
the average extension of the voids in the distribution of galaxies.
The average value of voids diameter can be derived from the following
proportion :
\begin{equation}
\frac {A_{max}}{\overline{A}}=
\frac {A_{max}^{obs}}{\overline{A}^{obs}}
\quad ,
\end{equation}
where the left hand side refers to the maximum and average 
value of the simulated  cross-sectional area named $V_p(2,3)$
(their numerical values are visible in the captions of 
Figure~\ref{2D_area_gamma})
and the right  hand side refers to the same quantities on the 
CFA2 slices.
A value  for the average observed diameter
,$\overline{D^{obs}}$  ,
is easily found
from the  previous proportion :
\begin{equation}
\overline{D^{obs}} \approx  0.6  {D_{max}^{obs}} = 2700 \frac{Km}{sec} 
= 27~Mpc 
\quad ,
\label{dobserved}
\end {equation} 
where $D_{max}^{obs}=4500~\frac{Km}{sec}$  corresponds to the 
extension of the maximum void visible on the CFA2 slices.
The half value of $\overline{D^{obs}}$ can be equated with  
equation~\ref{rprimeval} that gives the radius of the explosion from 
primeval galaxies and  the following is obtained:
\begin{equation}
{{\frac {{\it E_{64}}\,{{\it t_9}}^{2}}{{\it n_{-7}}}}}
= 1.47 
\quad . 
\label {primeval_gal}
\end {equation}
This  relationship   regulates  the three basic physical parameters
involved in  the explosions of primeval galaxies.

\begin{figure}
\begin{center}
\psfig{file=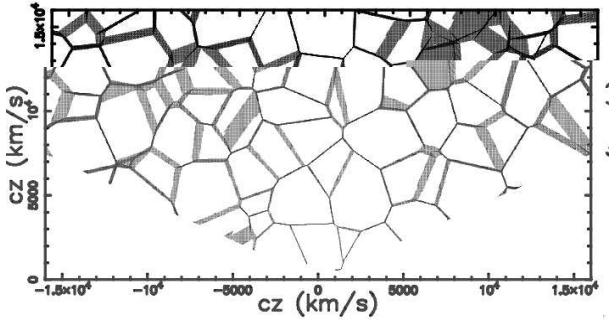,width=8cm} 
\end {center}
\caption {
Polar plot
of the  little cubes belonging to a
slice   $130^{\circ}$~long  and $6^{\circ}$
wide.
Parameters  as in Figure~\ref{cut_middle}.
}
          \label{cfaslices}%
    \end{figure}

The results of the simulation can be represented by  a
 slice
similar to that observed (a strip of $6^{\circ}$ wide and about
$130^{\circ}$ long) , see Figure~\ref{cfaslices}.

%begin figure cfaslices
% end figure cfaslices
%inserire%

\begin{figure}
\begin{center}
\psfig{file=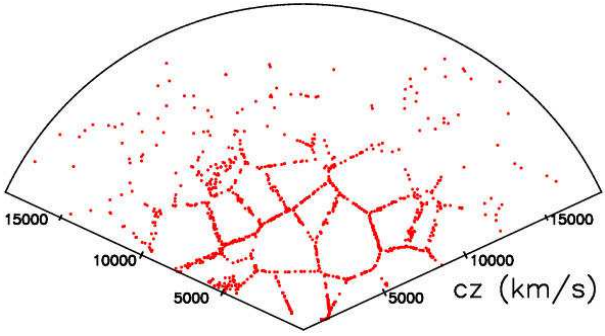,width=8cm} 
\end {center}
\caption 
{
Polar plot of the little cubes (red points) 
when  the "scaling" algorithm
is applied.
NBIN=15 and other parameters  as in Figure~\ref{cfaslices}~.
}
          \label{simu_mia}%
    \end{figure}

\begin{figure}
\begin{center}
\psfig{file=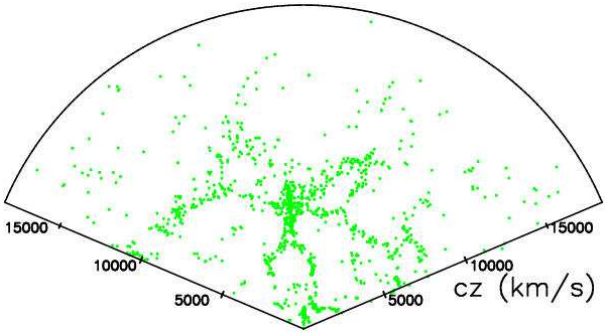,width=8cm} 
\end {center}
\caption { Polar plot of   the   real galaxies (green  points)
belonging to the second CFA2 redshift catalogue.}
          \label{simu_cfa}%
    \end{figure}

\label{algorithm}
For a more accurate  confrontation between simulation and
observations the effect due to the distribution in luminosity
should be introduced .
 Here
conversely a "scaling" algorithm is adopted
 that is now
summarised~:
\begin{enumerate}
\item
 The field of velocity of the observed sample is divided
 in NBIN intervals equally spaced.
\item
  In each of these NBIN intervals the number of galaxies NGAL(j) ,
   (j identifies  the selected interval) , is computed.
\item
  The field of velocity of the simulated little cubes
  belonging to the faces is sampled  as in point (1).
\item
  In each interval of the simulated field of velocity
  NGAL(j) elements are randomly selected .
\item
  At the end of this process the number of the little
  cubes  belonging
  to the faces  equalises the number and the scaling
  of the observed galaxies.
\end{enumerate}
A typical polar plot  once the "scaling" algorithm is  implemented
is  reported in Figure~\ref{simu_mia}; 
it should be compared 
with the observations  , see Figure~\ref{simu_cfa}. 

%begin figure simu_mia
% end figure simu_mia

%begin figure simu_cfa
% end figure simu_cfa

\subsubsection{The density of voids}

\label{voidsdensity}
The density of seeds expressed in physical units , $\rho_N$,
is the inverse
of the physical averaged volume, $\rho_N = \frac {1} {u_V^p}$,
and therefore
\begin{equation}
u^p_{A(2,3)}  =
(u^p_V)^{2/3}
=
(\frac {1}{\rho_N}) ^{2/3}
\quad  .
\end{equation}
At the same time the sectional area
will be characterised by a maximum physical area
, $A^p_{max}$ expressed in physical units,
\begin{equation}
A^p_{max} =
C_{Amax} \times  u^p_{A(2,3)}
\quad ,
\end{equation}
and the following is easily found 
\begin{equation}
\rho_N = ( \frac { C_{Amax}}{A^p_{max}})^{3/2}
\quad .
\end {equation}
We are now  ready to compute the number of sources in a sphere
of radius $R_{obs}=16000~Km/sec$~, the same radius
that characterises the CFA2 slices.
The number of voids/seeds
in the sphere turns out to be
\begin{equation}
N = \frac {1.7~10^{13} (C_{Amax})^{3/2} }
          { (A^{obs}_{max})^{3/2}}
\quad ,
\end{equation}
where  $A^p_{max}$ was identified with  $A^{obs}_{max}$.
Inserting   the dimension of the maximum void as deduced in
\S.~\ref{CFA2},
 we obtain $A^{obs}_{max}=1.59~10^7 (Km/sec)^2$.
We now have an expression for the number of seeds in the sphere
that characterises the CFA2
\begin{equation}
N=268  (C_{Amax})^{3/2} = 827
\quad .
\end{equation}

\section{Conclusions}

The Voronoi diagrams explain and characterise the voids in the
spatial distribution of galaxies. 
The characteristics of a typical
cell in a two--dimensional section of a 3D Voronoi diagram can be
compared with those connected with the 
voids in the distribution of galaxies. 
The  following items turn out to be useful to the astronomer once
$A^{obs}_{max}$~, the maximum area connected with a void, is
derived from the astronomical  slices:
\begin{itemize}
\item The averaged value of the  voids in the distribution of galaxies
      should be  $2741 \pm 210$
      Km/sec across.
\item The pdf of the area of the  voids in the distribution of galaxies 
      should be a gamma--variate with argument
      1.9.
\item The expected averaged value of the sides of the
      irregular polygons  that characterises the 
      voids in the distribution of galaxies
      should 5 . 
\end{itemize}

Some important key questions are tentatively addressed
to  the astronomical community:
\begin{itemize}
\item The maximum area connected with a void should be
      derived  with a great accuracy in the various slices.

\item The algorithms of describing polygonal
      voids from the astronomical observations should be
      developed in order to test the suggested averaged
      number of sides, 5 ,
      as predicted from the Voronoi diagrams.
\item The  pdf of the area   connected with
       the voids should be tentatively
       computed in order
      to test the goodness of the
      $V_p(2,3)$ predictions.
\end{itemize}

The Voronoi diagrams allow also to reformulate the theory
of the primordial explosions because 
\begin{itemize}
\item 
The average diameter of voids between galaxies is function
of three parameters~:  time , density and energy ,
see formula~\ref{primeval_gal}.
\item 
The galaxies are originated where the primordial shells 
meet , the faces of the irregular polyhedra.
\item
The correlation length for galaxies can be identified 
with the face's thickness that is approximately 
$\frac{1}{6}$  the radius of the expanding   shell.
\end {itemize}

\begin {acknowledgements}

       I thank
       the Smithsonian Astrophysical Observatory and
       John Huchra  for the small catalog available
       from the web at
       http://cfa-www.harvard.edu/~huchra/zcat/.
       I thank  the referee for useful  suggestions.

\end  {acknowledgements}

%\bibliography{biblio}

\label{lastpage}
\end {document}